\documentclass[article,showpacs,pra,reprint,longbibliography,superscriptaddress]{revtex4-1}

\usepackage[colorlinks=true,linkcolor=blue,citecolor=blue,urlcolor=blue]{hyperref}

\usepackage{setspace} 
\usepackage{graphicx}
\usepackage{amsmath}
\usepackage{color}
\usepackage{amsmath}
\usepackage{amssymb}
\usepackage{verbatim}
\usepackage{latexsym}
\usepackage{enumerate} 
\usepackage{bm} 
\usepackage{multirow}
\usepackage{booktabs}
\usepackage[caption=false]{subfig}
\usepackage{floatrow}
\begin{document}

\title{Finite temperature optoelectronic properties of BAs from first principles}

\author{Ivona Bravi\'{c}} \email{ib376@cam.ac.uk}
\author{Bartomeu Monserrat} \email{bm418@cam.ac.uk}
\affiliation{TCM Group, Cavendish Laboratory, University of Cambridge, J.\,J.\,Thomson Avenue, Cambridge CB3 0HE, United Kingdom}

\date{\today}

\begin{abstract} 

The high thermal conductivity of boron arsenide (BAs) makes it a promising material for optoelectronic applications in which thermal management is central. 
In this work, we study the finite temperature optoelectronic properties of BAs by considering both electron-phonon coupling and thermal expansion. The inclusion of electron-phonon coupling proves imperative to capture the temperature dependence of the optoelectronic properties of the material, while thermal expansion makes a negligible contribution due to the highly covalent bonding character of BAs. We predict that with increasing temperature the optical absorption onset is subject to a red shift, the absorption peaks become smoother, and the phonon-assisted absorption at energies below those of the optical gap has a coefficient that lies in the range $10^{-3}$--$10^{-4}$\,cm$^{-1}$. We also show that good agreement with the measured indirect band gap of BAs is only obtained if exact exchange, electron-phonon coupling, and spin-orbit coupling effects are all included in the calculations.  
\end{abstract}

\maketitle

\section{Introduction} \label{Intro}



Semiconductors from the III--V main group, such as gallium arsenide (GaAs), possess a variety of properties which make them suitable candidates to compete with silicon in optoelectronic applications. The III--V \textit{zinc-blende} type compound boron arsenide (BAs) had until recently not been considered a suitable competitor in this context due to the inability of synthesizing high quality samples. However, the recent successful synthesis of a highly pure sample of cubic BAs, and especially the prediction \cite{Broido2013_Abinitio_thermal_transport_BAs,Broido2016_thermal_transport_vacancies, Slack1973_high_thermal_cond,Tian2019_BAs_review} and subsequent measurement of an extraordinarily high room temperature thermal conductivity in the range of $900-1300$\,W/mK \cite{Science_BAs_1,Science_BAs_2,Science_BAs_3}  (second only to diamond), have placed BAs at the core of optoelectronics research. 

BAs could prove particularly useful for thermal management, which typically poses a big challenge for operating optoelectronic devices. Additionally, its low density, large resistivity, low effective carrier mass \cite{Kioupakis_BAs_GW,Lui2018_Carriers_BAs,2014_vibrations_BAs}, the possibility of making a perfect alloying system with GaAs \cite{Zunger_Band_BAs-alloys}, chemical resistance towards decomposition and dissolution at ambient conditions \cite{Robert1960_Chem_props_BAs}, and an electronic configuration which is isoelectronic to that of Si, open doors to a multitude of novel optoelectronic applications for BAs \cite{Franklin2016_zbBAs_calc,Ge2019_optics_BAs,Kioupakis2019_BAs_GW,Kioupakis_BAs_defects,Lindsay2013_BAs_vs_Diamond}, including photovoltaics and photoelectrochemical water splitting \cite{Wang2012BAs_BG_exp}.

The optical properties of BAs are dominated by an indirect band gap at $1.46$\,eV \cite{Wang2012BAs_BG_exp} which determines the absorption onset of the optical spectrum. Momentum conservation dictates that indirect optical transitions are mediated by phonons and therefore the absorption across the indirect gap is a relatively weak second order process \cite{Williams1951_QZPE,Lax1952_Franck-condon_crystals,Hall1954_HBB_indirect_absorption}. Stronger absorption is expected starting at the energy of the minimum direct gap, which in BAs occurs at the $\Gamma$-point. The electronic structure of BAs is also influenced by the spin-orbit interaction, which lifts the triple degeneracy of the valence band maximum, and by electron-electron correlation effects, which are required to capture the correct magnitude of the band gap \cite{Kioupakis_BAs_GW, Levy1996_BG_Problem,MUSCAT2001_BG_Hybrids,Richard2005_HSE_BG}. 
Furthermore, temperature changes have been shown to significantly modify the band gap \cite{Giustino2010_EPC_diamond,elph_Si_nano_Bester2013,Monserrat2014_C_Si_bg,Antonius2014_MB_effects_ZP_BG,qe_yambo_comparison} and absorption spectrum \cite{Kioupakis2010_indirect_abs_nitrides,Cannuccia2011_ZPQM_Optics,Noffsinger2012_Si_phonon,Giustino2014_EPC+optics,Giustino2015_EPC+optics2,Tomeu2018_BaSnO3,Morris2018_abs,Noel2018_abs,Kresse2018_K-edge_hbn} of multiple semiconductors.

In this work, we use first principles methods to study the role of temperature on the optoelectronic properties of BAs by including both electron-phonon interactions and thermal expansion. Our main observations are: (i) the inclusion of electron-phonon coupling, spin-orbit coupling, and exact exchange are all essential to reproduce the experimentally measured band gap, (ii) the absorption spectrum smoothens with increasing temperature, (iii) the indirect band gap undergoes a red shift of about $110$\,meV due to quantum zero point motion, and the absorption onset a red shift of a further $80$\,meV with increasing temperature from $0$\,K to $400$\,K, (iv) the electron-phonon coupling strength is weakly dependent on spin-orbit interactions but changes by about $20$\% when exact exchange is included, (v) terms beyond the lowest order in the electron-phonon interaction make a significant contribution and increase the band gap correction by about $30$\%,  and (vi) thermal expansion makes a negligible contribution to the optoelectronic properties of BAs.

The rest of the paper is organized in the following fashion. In Sec.\,\ref{Equilibrium_props} we discuss the equilibrium properties of \textit{zinc-blende} BAs, differentiating between results using various exchange-correlation functionals, and discussing both electronic structure and lattice dynamics. These calculations serve as a prerequisite for Sec.\,\ref{Temperature_effects_GS}, where we investigate the finite temperature effects of the electronic structure driven by electron-phonon coupling and thermal expansion. In Sec.\,\ref{optics} we demonstrate the effect of electron-phonon coupling on the finite temperature optical absorption of BAs, and we summarize our findings in Sec.\,\ref{Summary}. 

\section{Equilibrium properties} \label{Equilibrium_props}

\subsection{Computational details}

All our first principles calculations are performed using density functional theory (DFT) \cite{HK1964_Electron-gas,Sham1965_SCF_EXC} with the projector augmented wave method \cite{Han2012_PAW,Joubert1999_PAW2} as it is implemented in {\sc vasp} \cite{Kresse1996_VASP,Kresse1996a_VASP2,Kresse_VASP3,Kresse_VASP4}. We employ an energy cutoff of $500$\,eV and a Brillouin zone (BZ) grid including $8\times8\times8$ $\mathbf{k}$-points for the primitive cell and commensurate grids for the supercells. We relax the volume until the stress is below $10^{-2}$\,GPa, while the internal atomic coordinates are fixed by symmetry. We compare the performance of two different exchange correlation functionals, namely the semilocal generalized gradient approximation of Perdew-Burke-Ernzerhof (PBE) \cite{Perdew1996_GGA} and the hybrid Heyd-Scuseria-Ernzerhof functional (HSE) \cite{Paier2006a_HSE,Paier2006_HSE_erratum}. The spin-orbit interaction is included perturbatively using the second variational method \cite{Koelling1977_SOC}.

For the lattice dynamics calculations, we employ the finite displacement method \cite{Kunc1982_finite_displacement} combined with nondiagonal supercells \cite{Tomeu2015_nondiag_supercells} to construct the matrix of force constants, which is then Fourier transformed to the dynamical matrix and diagonalized to obtain the vibrational frequencies and eigenvectors. We obtain converged results using a $6\times6\times6$ coarse \textbf{q}-point grid used as a starting point for the Fourier interpolation to a finer grid along high symmetry lines to construct phonon dispersions and to a finer stochastic grid to calculate the vibrational contribution to the energy. We only use the PBE functional for the lattice dynamics calculations, and avoid using the HSE functional due to the increased computational cost associated with the incorporation of exact exchange. 

\begin{table*}
  \setlength{\tabcolsep}{6pt} 
  \caption{Calculated structural and electronic properties of \textit{zinc-blende} BAs employing the PBE and HSE exchange correlation functionals with and without incorporation of spin-orbit coupling. We include the calculated cubic static lattice parameter with a comparison to the experimental value from Ref.\,\cite{Lattice_constant_BAs}, the direct optical band gap $\mathrm{E_{g}^{direct}}$ at the $\Gamma$-point, the indirect minimum band gap $\mathrm{E_{g}^{ind}}$ as the difference of the conduction band minimum and the valence band maximum with a comparison to the experimentally reported value from Ref.\,\cite{Wang2012BAs_BG_exp}, and the spin-orbit gap $\mathrm{E_{g}^{SO}}$ at the valence band maximum of the $\Gamma$-point.}
  \label{Benchmark}
  \begin{ruledtabular}
  \begin{tabular}{lccccc}
  & Lattice parameter \textit{a} ($\mathrm{\AA}$)& $\mathrm{E_{g}^{direct}}$ (eV) & $\mathrm{E_{g}^{ind}}$ (eV) & $\mathrm{E_{g}^{SO}}$ (eV) \\ 
  \hline
PBE & 4.817&3.242 & 1.197& --& \\
PBE + SOC & 4.817&3.036 & 1.132 & 0.206\\
HSE &4.770&4.097&1.768&-- \\
HSE + SOC & 4.770&3.905&1.717&0.192 \\ 
Experiment & 4.777&-- &1.46& -- \\
\end{tabular}
\end{ruledtabular}
\end{table*}

\subsection{Structure}\label{Equilibrium_structure}

We consider the cubic $F\overline{4}3m$ structure of BAs with two atoms in the primitive cell and with an experimentally reported lattice parameter of $a=4.777$\,\AA\@ \cite{Lattice_constant_BAs}. In order to find the equilibrium geometry we minimize the energy with respect to the cubic lattice parameter \textit{a}. The results for the different functionals listed in the previous section are shown in Table \ref{Benchmark}. We find an equilibrium geometry for PBE in which the lattice parameter overestimates the experimental one by about 0.8\,\%. For HSE, we obtain a smaller volume that only underestimates the experimental value by 0.2\,\%, following the standard trend reported for hybrid functionals \cite{Scuseria2012_Benchmark_Lattice}. The results are independent of whether the spin-orbit interaction is included or not.

 
\subsection{Bandstructure}\label{bandstructure}

\begin{figure}	
	\centering
	\includegraphics[scale=0.38]{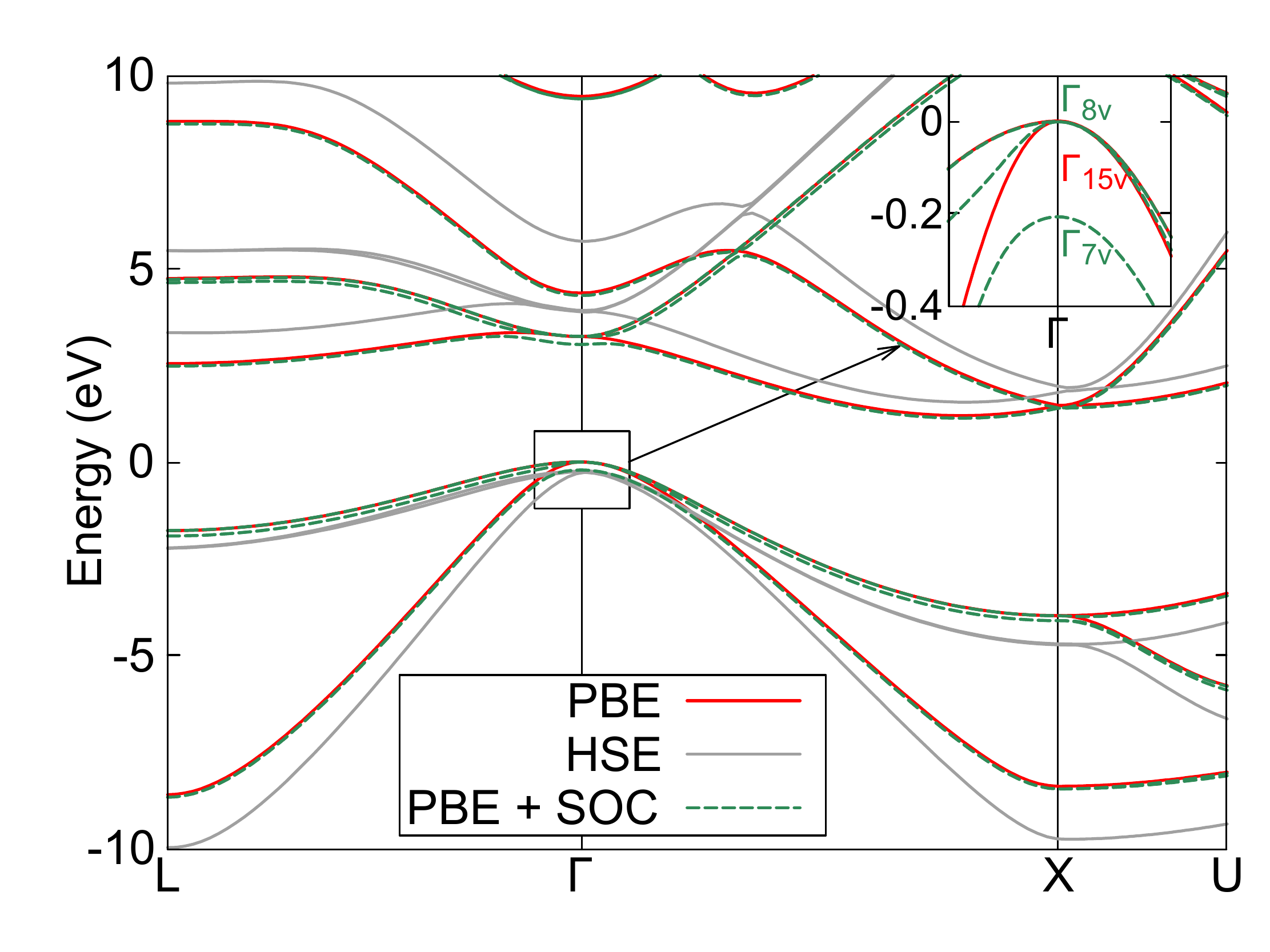}
	\caption{Electronic band structure of BAs calculated at the PBE and HSE level of theory. The band structure for PBE is plotted both with and without spin-orbit effects and we show the relevant band representations of the VBM with and without spin-orbit coupling.}
	\label{fig:bandstructure}
\end{figure}

In Fig. \ref{fig:bandstructure} we show the bandstructure of BAs calculated using the PBE and the HSE functionals. The valence bands are dominated by both boron and arsenic $p$-type orbitals with the valence band maximum at the  six-fold degenerate Gamma point $\Gamma$ (0,0,0). The conduction band at the $\Gamma$-point is also a $p$-type state constituted of 33\,\% B 2\textit{p}, 61\,\% As 4\textit{p} and 6\,\% As 3\textit{d}$_{xz}$ orbitals. The conduction band minimum is located on the path between the $\Gamma$ and the X-point at the fractional $\mathbf{k}$-point (0.39, 0, 0.39) and has significant additional \textit{s}-type contributions from both boron and arsenic. 

 The minimum band gap is of indirect nature and has a value of $1.19$\,eV calculated with PBE and of $1.76$\,eV calculated with HSE. By comparison, the experimental estimates of the minimum band gap lie at around $1.46$\,eV as measured at $300$\,K \cite{Wang2012BAs_BG_exp}. The minimum direct band gap occurs at the $\Gamma$-point and has a value of $3.24$\,eV for PBE and $4.09$\,eV for HSE with no experimental estimates available for comparison. We provide a list of the minimum indirect and direct gaps calculated with a range of exchange correlation functionals in Table \ref{Benchmark}.

 The bandstructure of BAs calculated with the HSE functional and shown in Fig.\,\ref{fig:bandstructure} shows two main distinguishing features when compared to the result without incorporating exact exchange. Firstly, there is an overall shift of the conduction bands towards higher band energy values while the valence bands are stabilized. Secondly, the magnitude of the energy shift is different for individual $\mathbf{k}$-points and bands. This change in dispersiveness can be quantified by the changes in the direct and indirect band gap values, the optical band gap increases by $0.85$\,eV while the indirect band gap changes by only $0.57$\,eV. 

When spin-orbit coupling is incorporated into the ground state bandstructure calculation we observe a lift  of the degeneracy of both the valence and the conduction bands, as expected in \textit{zinc blende}-type structures \cite{Parmenter1955_zincblende_symmetries}. The valence band maximum six-fold degenerate $\Gamma_{15v}$ state is separated into a four-fold degenerate $\Gamma_{8v}$ band (with $j=3/2$) and a two-fold degenerate $\Gamma_{7v}$ band with a spin-orbit (SO) gap of about $200$\,meV (see Table \ref{Benchmark}). 

In Sec.\,\ref{Temperature_effects_GS} below we will show the finite temperature effects on the band gap using predominately the PBE functional, but we also consider the effect of electron-electron correlation on the electron-phonon coupling strength by cross-checking with the HSE functional.

\subsection{Lattice dynamics}\label{Lattice_dynamics}

\begin{figure}
\centering
\includegraphics[scale=0.37]{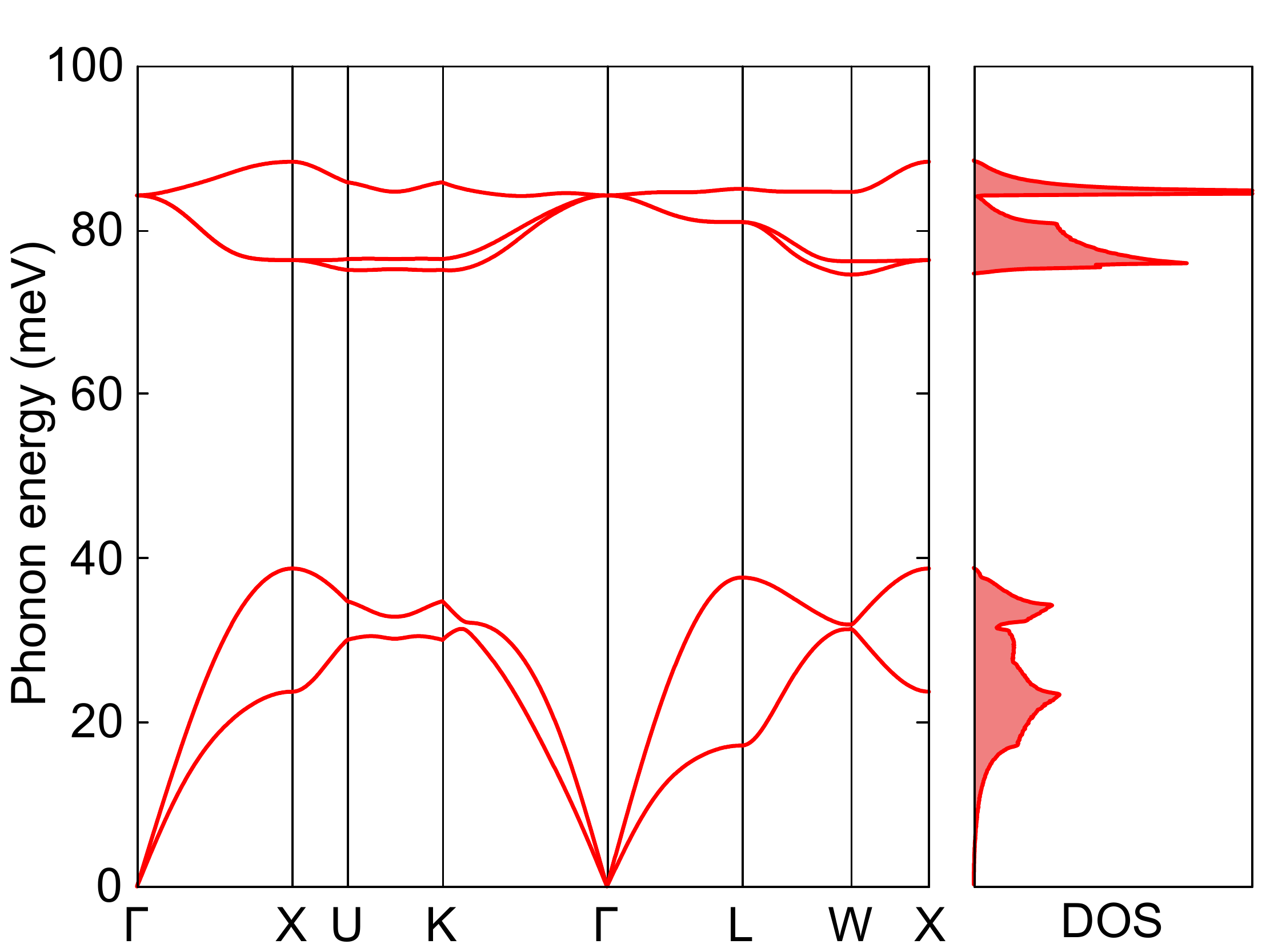}
\caption{Phonon dispersion (left) and density of states (right), calculated with $6\times6\times6$ $\mathbf{q}$-point coarse grid.}	
\label{fig:phonon_dispersion}
\end{figure}

We show the phonon dispersion curve in Fig.\,\ref{fig:phonon_dispersion} calculated using the PBE functional along a high symmetry path of the BZ.
The phonon dispersion does not exhibit any imaginary modes, indicating that the structure under consideration is dynamically stable. We note that some systems exhibit a dynamical instability when using the PBE functional that disappears when a range-separated hybrid functional is used \cite{Tomeu2018_BaSnO3,Lazzeri_electron-corellation+EPC}, and in those cases it becomes necessary to describe the lattice dynamics at the higher level of theory. This latter case is not applicable in our system, and we therefore only use the PBE functional for the lattice dynamics calculations due to the large computational cost associated with the HSE functional calculations.


Experimental work has shown that there is no LO-TO splitting in cubic BAs \cite{Chu2014_BAs_LOTO}, a feature that is correctly reproduced in our study since the calculated value of LO-TO splitting lies in the sub-meV region. This weak LO-TO splitting is due to the high covalent character of the compound, reflected in the very similar values of the high and low frequency dielectric constants of $\epsilon_{\infty}=9.717$ and $\epsilon_{0}=10.175$, as well as in the Born effective charge of $|Z^{*}| = 0.46$. 
The unusually high covalency of BAs has also been highlighted in earlier theoretical studies \cite{Covalency_BAs}.

The phonon dispersion calculated including spin-orbit coupling is almost indistinguishable from the one reported in Fig.\,\ref{fig:phonon_dispersion}. As an example, the optical frequencies at the $\Gamma$-point differ by less than $0.6$\,meV. 

\section{Temperature dependence of the electronic structure}\label{Temperature_effects_GS}

We have thus far considered the ground state properties of BAs within the static lattice approximation. However, for the use of thermoelectrics in optoelectronic applications it becomes interesting to study the effects of temperature on the optical and electronic responses of the material. In this section we focus on the electronic structure and consider two effects that can give rise to temperature-induced changes of the band structure, namely electron-phonon coupling and thermal expansion. In both cases we will study how the indirect band gap changes with increasing temperature. \\

\subsection{Computational details} \label{sec:comp_det_elec}

At temperature $T$, the electron-phonon coupling renormalized electronic band $\epsilon_{n\mathbf{k}}$, labelled with the momentum $\mathbf{k}$ and band index $n$, is given by the expectation value: 
\begin{equation}
\epsilon_{n\mathbf{k}}(T)=\frac{1}{\mathcal{Z}}\sum_{\mathbf{s}}\langle\Phi_{\mathbf{s}}(\mathbf{u})|\epsilon_{n\mathbf{k}}(\mathbf{u})|\Phi_{\mathbf{s}}(\mathbf{u})\rangle e^{-E_{\mathbf{s}}/k_{\mathrm{B}}T}, \label{eq:elec_tdep}
\end{equation}
where $|\Phi_{\mathbf{s}}\rangle$ is the vibrational wave function in state $\mathbf{s}$ and with energy $E_{\mathbf{s}}$, evaluated within the harmonic approximation in this work, and $\mathcal{Z}=\sum_{\mathbf{s}}e^{-E_{\mathbf{s}}/k_{\mathrm{B}}T}$ is the partition function in which $k_{\mathrm{B}}$ is Boltzmann's constant. Equation\,(\ref{eq:elec_tdep}) shows an explicit dependence on $\mathbf{u}=\{u_{\nu\mathbf{q}}\}$, a collective coordinate for all the nuclei written in terms of normal modes of vibration with wavevector $\mathbf{q}$ and branch $\nu$. We evaluate Eq.\,(\ref{eq:elec_tdep}) using two approaches, the \textit{quadratic approximation} that relies on a low-order expansion of the dependence of $\epsilon_{n\mathbf{k}}$ on $\mathbf{u}$ \cite{Allen1976_Temp_Bands}, and the \textit{Monte Carlo approximation} that relies on a stochastic evaluation of Eq.\,(\ref{eq:elec_tdep}), that we accelerate using thermal lines \cite{Tomeu2016_TL}. The quadratic approximation can be combined with nondiagonal supercells \cite{Tomeu2015_nondiag_supercells} to converge the electron-phonon coupling contribution with respect to the number of phonon wavevectors used to sample the vibrational BZ at the expense of neglecting terms beyond second order in the expansion of $\epsilon_{n\mathbf{k}}$ in terms of $\mathbf{u}$. The Monte Carlo approximation is typically restricted to smaller simulation cells (coarser $\mathbf{q}$-point grid sampling) but includes the dependence of $\epsilon_{n\mathbf{k}}$ to all orders in $\mathbf{u}$. Using both approaches allows us to identify the most important contributions to the electron-phonon coupling effect. We refer the reader to Ref.\,\cite{Tomeu_EPC-review} for a recent review of these methods. 

We calculate the equilibrium volume of BAs at temperature $T$ by minimizing the Helmholtz free energy as a function of volume within the quasiharmonic approximation \cite{Dove2005}. The renormalization of electronic bands $\epsilon_{n\mathbf{k}}$ arising from thermal expansion is then evaluated using the appropriate volume at each temperature. Convergence tests show that the relative free energy difference between coarse grids of sizes $6\times6\times6$ and $8\times8\times8$ $\mathbf{q}$-points is below $2$\,meV/atom at $300$\,K. Accordingly, the results in Sec.\,\ref{Thermal_expansion} are obtained with a $6\times6\times6$ $\mathbf{q}$-point grid.

\subsection{Electron-phonon coupling} \label{sec:el-phonon-coupling}

\begin{figure}
\centering
	\sidesubfloat[]{\includegraphics[scale=0.28]{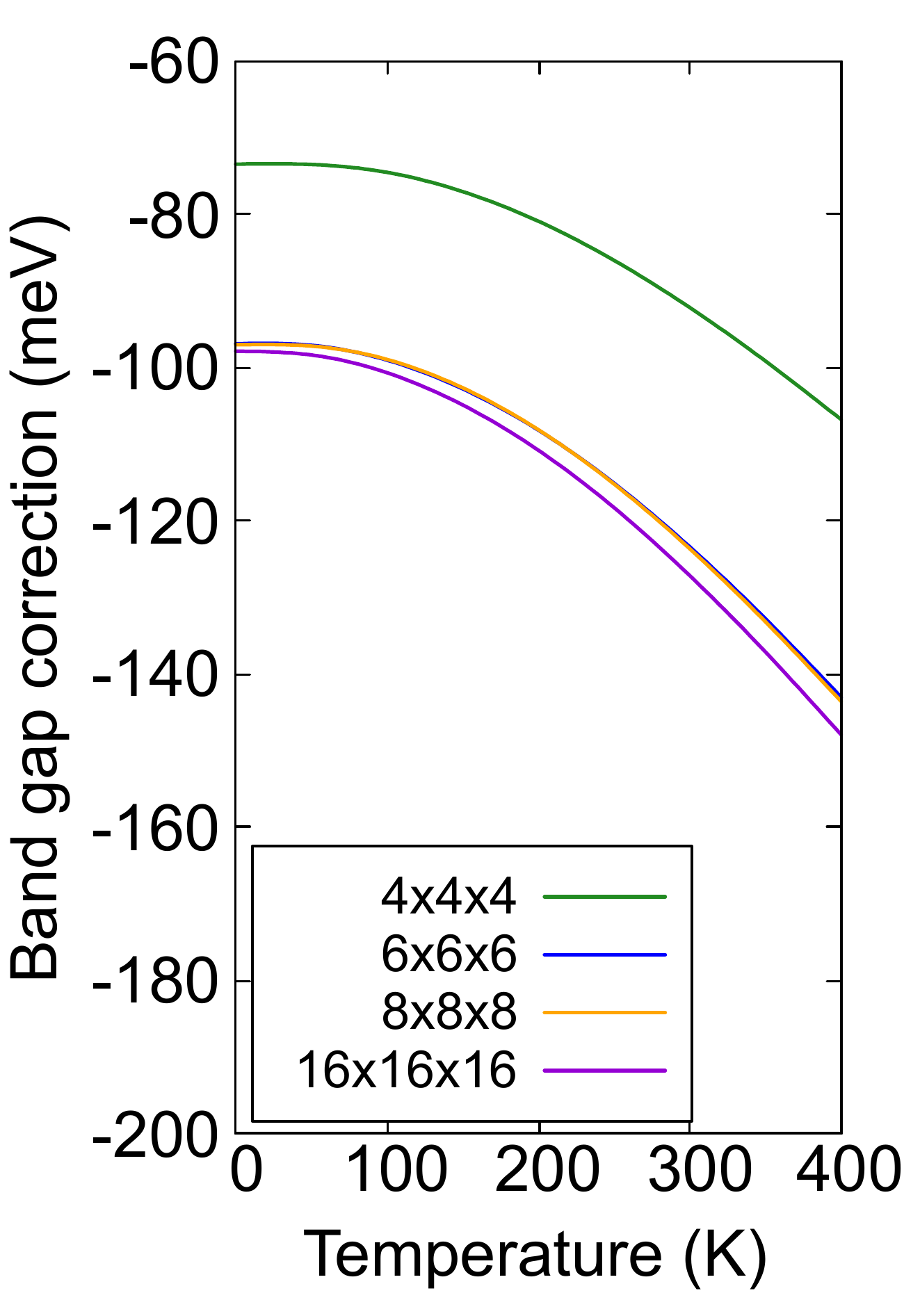}\label{fig:el-phonon-coupling-a}}
	\hfil
	\sidesubfloat[]{\includegraphics[scale=0.28]{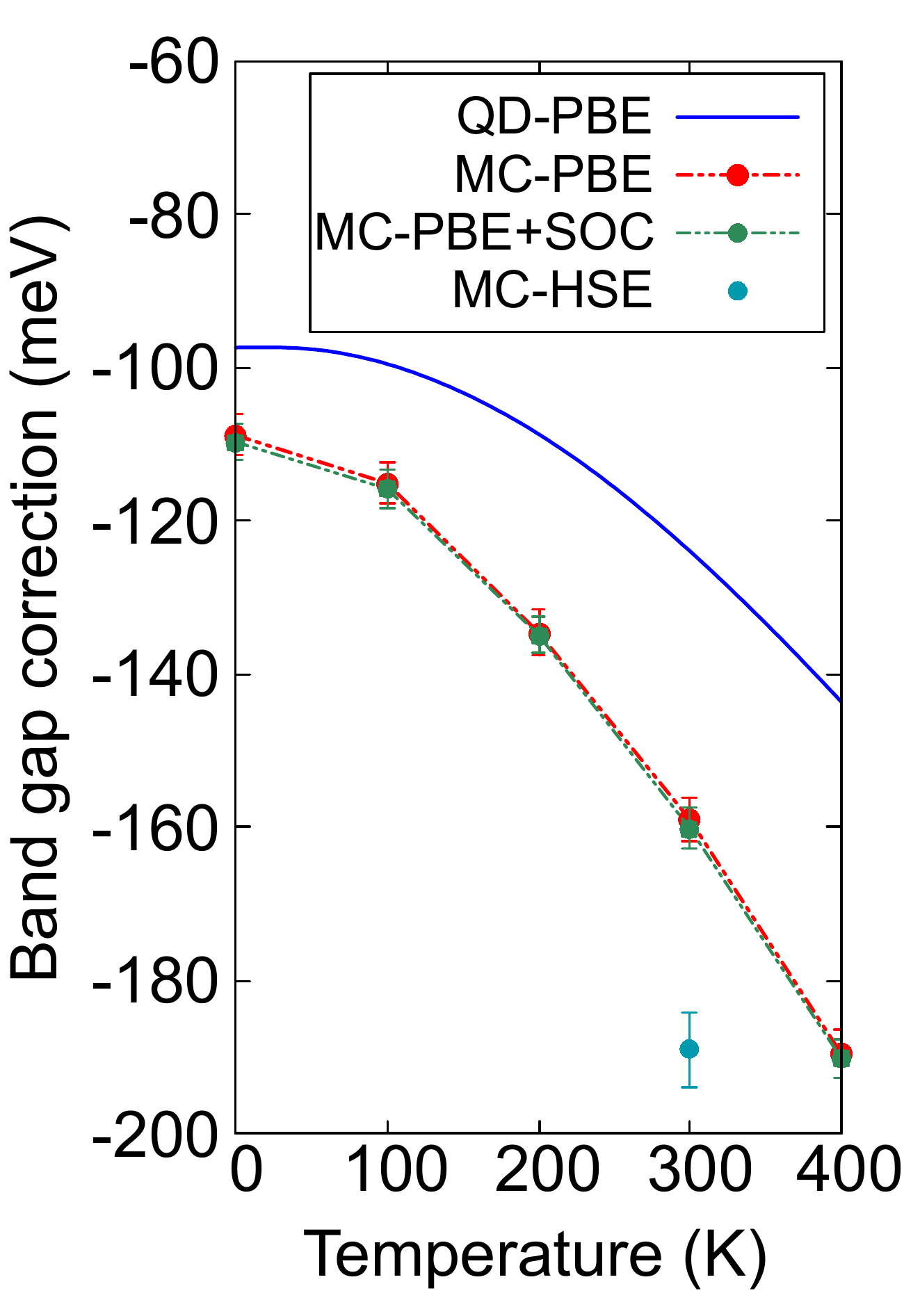}\label{fig:el-phonon-coupling-b}}
	\caption{(a) Temperature dependence of the indirect band gap correction  arising from electron-phonon coupling, computed with PBE, and evaluated within the quadratic approximation for different $\mathbf{q}$-point grid sizes. Note, that the band gap corrections for the $6\times6\times6$ and $8\times8\times8$ $\mathbf{q}$-point grids almost entirely overlap. (b) Comparison between the results for the temperature dependent band gap correction of the indirect band gap obtained with the quadratic approximation (blue line) and the MC sampling approach (red circles) using a $6\times6\times6$ $\mathbf{q}$-point grid in both approaches. Within the MC scheme, we also include spin-orbit effects (green circles), and use the HSE functional at $300$\,K only (blue circle).}
	\label{fig:el-phonon-coupling}
\end{figure}

We apply the quadratic approximation and the stochastic Monte Carlo approximation using the PBE functional with an energy cutoff of 500\,eV to compute the zero point and finite-temperature renormalization of the indirect band gap of BAs, with the VBM at the $\Gamma$-point and the CBM located at the $\mathbf{k}$-point $(0.39,0,0.39)$ in the BZ. For the quadratic approach we benchmark the temperature dependence of the indirect band gap evaluated using $\mathbf{q}$-point grids of sizes $4\times4\times4$, $6\times6\times6$, $8\times8\times8$ and $16\times16\times16$ at finite displacement amplitudes of $0.5u^{\mathrm{RMS}}$ where $u^{\mathrm{RMS}}=\sqrt{\langle u^2\rangle}$ for each phonon mode. The convergence of the quadratic approximation as a function of $\mathbf{q}$-point grid size is depicted in Fig.\,\ref{fig:el-phonon-coupling-a}, and indicates that calculations on a $6\times6\times6$ $\mathbf{q}$-point grid are within $5$\,meV of the reference at $400$\,K, with better convergence for lower temperatures.
 For the Monte Carlo approach $80$ different atomic configurations are sampled in a $6\times6\times6$ supercell for $0$\,K to $300$\,K and $100$ configurations for $400$\,K, both with and without spin-orbit coupling, with the results depicted in Fig.\,\ref{fig:el-phonon-coupling-b}.  

Figure \ref{fig:el-phonon-coupling} shows that temperature dependent electron-phonon coupling reduces the indirect band gap significantly, which is in agreement with the behaviour exhibited by most semiconducting materials \cite{Cardona2005_bandgap}. However, we observe significant discrepancies between the quadratic method and the Monte Carlo approach when it comes to the zero point as well as the temperature dependent band gap correction. The MC evaluation predicts a band gap renormalization due to zero point quantum motion which is up to $10$\,\% larger than for the quadratic evaluation. At $400$\,K the MC calculations predict a band gap change which is $32$\,\% larger than for the quadratic approximation. This difference can be attributed to the higher-order phonon contributions to the electron-phonon coupling which are included in the Monte Carlo scheme but not in the quadratic approximation. The latter relies on the truncation of an expansion of $\epsilon_{n\mathbf{k}}(\mathbf{u})$ in terms of powers of $\mathbf{u}$ to second order, and thus neglects any higher-order contributions arising from non-parabolicity. This feature suggests that in BAs higher-order electron-phonon interaction terms play a significant role especially at higher temperatures.

For certain types of materials such as metal halide perovskites \cite{Tomeu2016_EPC_MAPbI3} or superconducting Pb \cite{Heid2010_SOC_EPC_superconductors} it has been shown that spin-orbit effects can modify the electron-phonon coupling strength significantly. Therefore, we also consider the interplay between spin-orbit coupling and electron-phonon coupling in BAs, which is reflected both in the size of the minimum indirect band gap and in the spin-orbit gap of the VBM at the $\Gamma$-point between the $\Gamma_{8v}$ and $\Gamma_{7v}$ states (see Fig.\,\ref{fig:bandstructure}). For the minimum band gap, Fig.\,\ref{fig:el-phonon-coupling-b} shows that the temperature driven correction including spin-orbit coupling effects lies within the statistical uncertainty of the corresponding correction without consideration of spin-orbit coupling. For the spin-orbit gap at $\Gamma$, the static lattice value of $\mathrm{E_{g}^{SO}}=206$\,meV is reduced by only $3$\,meV due to zero point quantum motion and decreases by a further $6\pm3$\,meV with increasing the temperature from $0$\,K to $400$\,K. Given the small changes induced by the inclusion of spin-orbit coupling in the strength of electron-phonon coupling, we will neglect spin-orbit effects when calculating the finite temperature optical properties in Sec. \ref{optics}.


Finally, we quantify the significance of the neglected electron-electron correlation on the electron-phonon coupling strength by including exact exchange via the HSE functional. Due to the computational expense associated with the HSE functional, our calculations are restricted to a $4\times4\times4$ supercell, and the results are compared to those of a $4\times4\times4$ supercell using the PBE functional in Table \ref{tab:indirect_PBE_vs_HSE}. At $300$\,K, the band gap correction of $-142$\,meV for PBE is enhanced to $-163$\,meV for HSE, thus indicating the importance of exact exchange on the strength of electron-phonon coupling. For both PBE and HSE band gap corrections, the shift in the individual VBM and CBM are comparable in magnitude, but with opposite signs. To obtain an accurate estimate of the band gap at $300$\,K, we note that the change in the band gap correction from PBE to HSE using a $3\times3\times3$ supercell is of $18\pm9$\,meV, which compared to the corresponding change of $21\pm8$\,meV for the $4\times4\times4$ supercell suggests that the effect of exact exchange on the electron-phonon coupling strength across the vibrational Brillouin zone is relatively uniform. This allows us to extrapolate the band gap value by using the difference between the converged $6\times6\times6$ PBE calculations and the unconverged $4\times4\times4$ PBE calculations as a finite size correction to the unconverged $4\times4\times4$ HSE calculations. The resulting HSE band gap at $300$\,K is $1.576$\,eV, as reported in Table \ref{tab:indirect_PBE_vs_HSE}. If we further account for the band gap correction due to spin-orbit coupling at the stattic lattice level obtained for HSE (see Table \ref{Benchmark}), we get a final value for the $300$\,K band gap of BAs of $1.52$\,eV, which is only $80$\,meV above the measured indirect band gap \cite{Wang2012BAs_BG_exp}.

Despite the increase of about $19$\% in the electron-phonon coupling strength in going from PBE to HSE, the overall band gap correction only changes by about $40$\,meV between these two levels of theory. Therefore, in Sec.\,\ref{optics} below we evaluate phonon-assisted optical aborption neglecting the contribution of exact exchange to the electron-phonon coupling strength, and only considering its larger effect on the blue shift of the static indirect band gap reported in Table \ref{Benchmark}.

\begin{table}
 \setlength{\tabcolsep}{6pt} 
 \caption{Indirect band gap of BAs within the static lattice approximation and including electron-phonon coupling effects at $300$\,K. We show the results using a $4\times4\times4$ supercell for both PBE and HSE, and converged results using a $6\times6\times6$ supercell for PBE. We also show a system size corrected HSE band gap (see text for details) and the results including spin-orbit coupling. The experimental band gap is taken from Ref.\,\cite{Wang2012BAs_BG_exp}.}
  \label{tab:indirect_PBE_vs_HSE}
  \begin{ruledtabular}
  \begin{tabular}{lcccc}
  & $\mathrm{E_{g}^{static}}$ & $\mathrm{E_{g}^{300K}}$& $\Delta\mathrm{E_{g}^{static-300K}}$\\[0.1cm]
& eV & eV & meV  \\ 
  \hline
PBE ($4\times4\times4$) & $1.199$ &  $1.057(6)$ & $-142(6)$ \\
PBE ($6\times6\times6$) & $1.197$ &  $1.038(5)$ & $-159(5)$ \\
HSE ($4\times4\times4$) & $1.768$ &  $1.605(5)$ & $-163(5)$ \\
HSE (corrected) & $1.765 $ & $1.576(5)$ & $-189(5)$ \\
HSE (corr.) + SOC & $1.714$ & $1.525(5)$ & $-189(5)$ \\
Experiment & --& $1.46$ & --\\
\end{tabular}
\end{ruledtabular}
\end{table}

\subsection{Thermal expansion}\label{Thermal_expansion}

The last effect we consider for the description of the finite temperature optoelectronic properties of BAs is the role of thermal expansion. In optoelectronic devices, especially when alloyed with other semiconducting materials, it is undesirable to observe large temperature driven volume changes of the system under operation. Therefore it is necessary to account for the volume change upon temperature increase and the influence on the absorption wavelength.

\begin{figure}
	\centering
	\sidesubfloat[]{\includegraphics[scale=0.31]{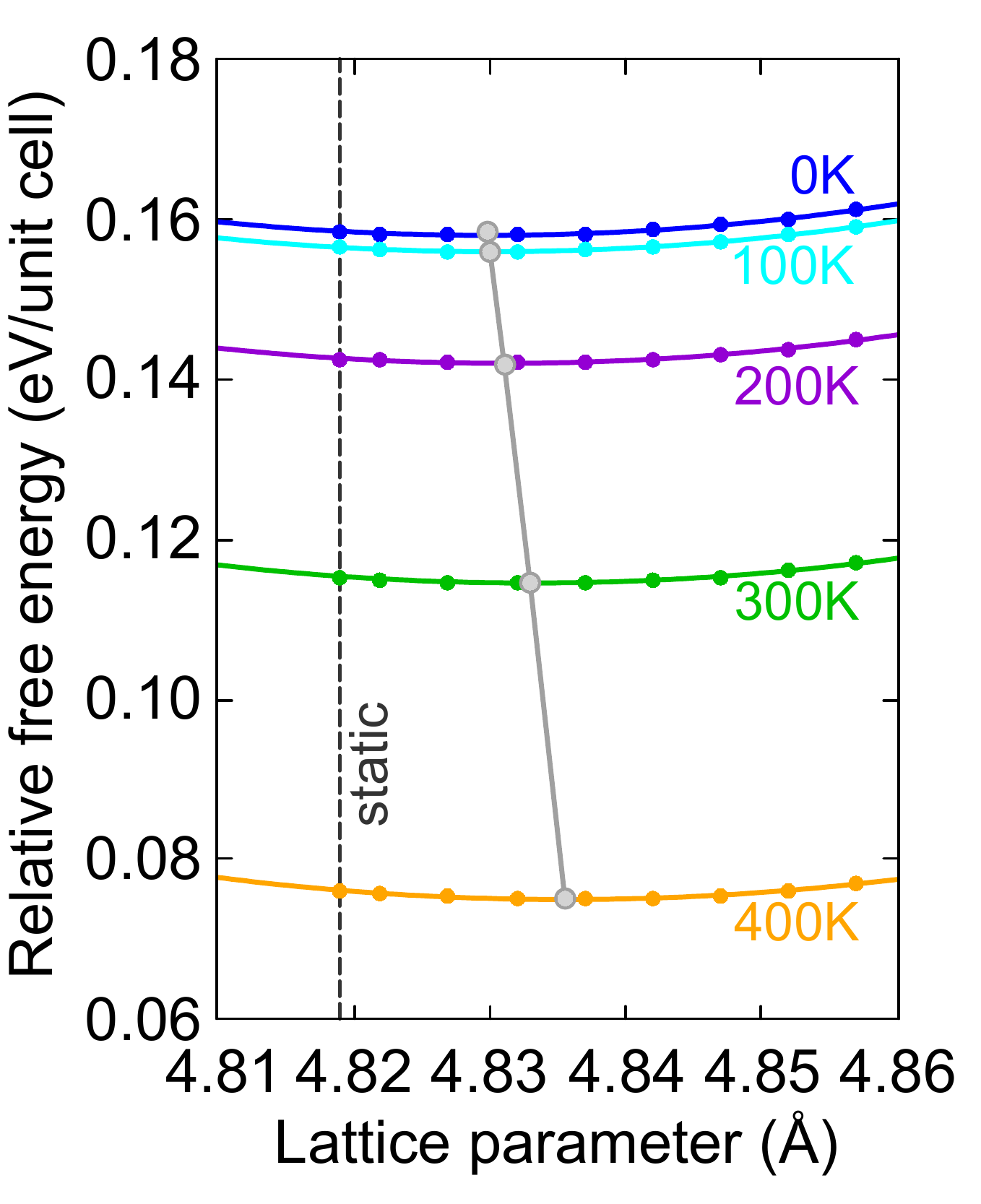}\label{fig:thermal-expansion_gap_a}}
	\hfil
	\sidesubfloat[]{\includegraphics[scale=0.31]{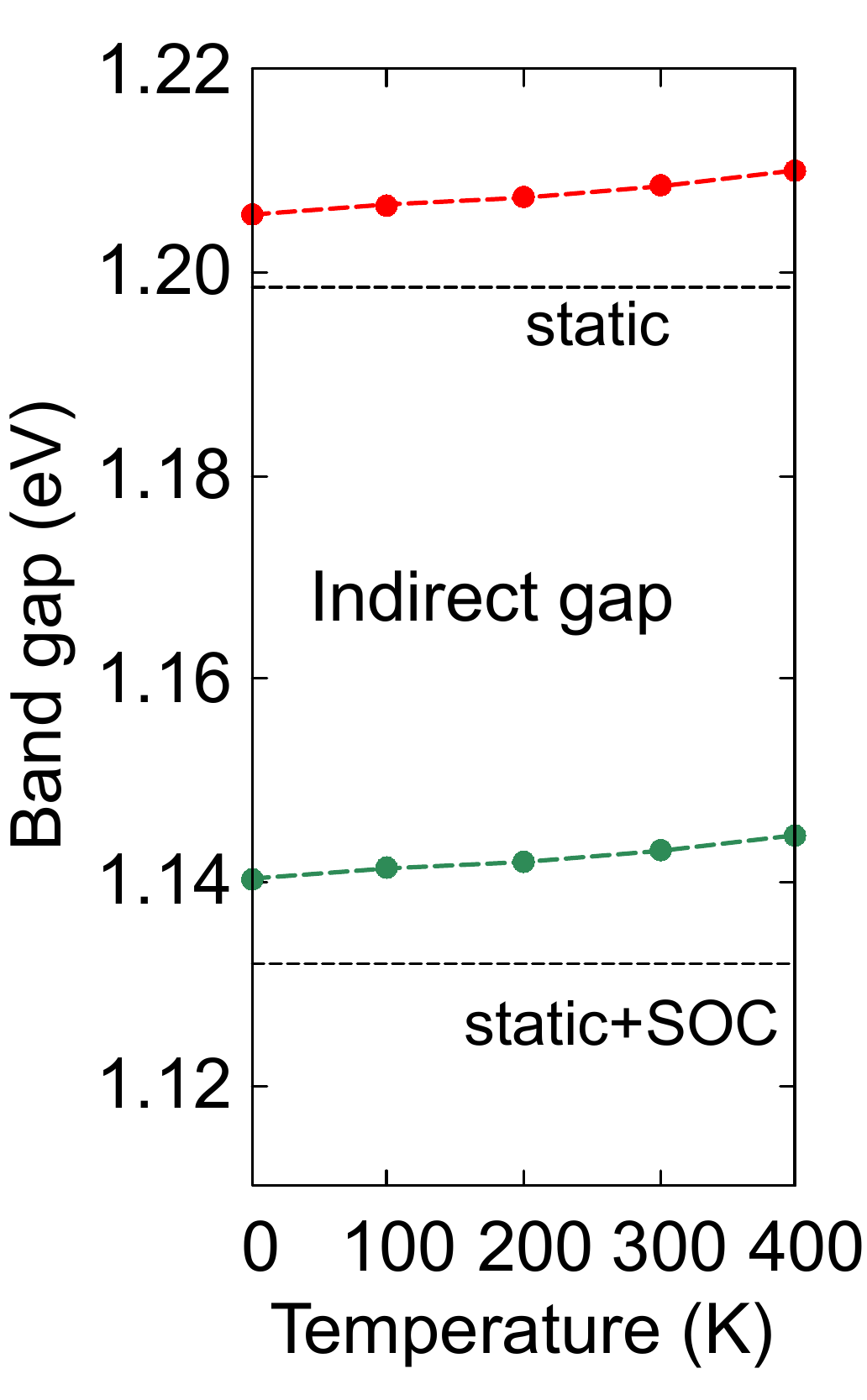}\label{fig:thermal-expansion_gap_b}}
	\caption{Thermal expansion of BAs. (a) Relative free energy as a function of the cubic lattice parameter for five distinct temperatures between $0$\,K and $400$\,K. The grey circles indicate the minima of quadratically fitted free energy curves at each given temperature. (b) Temperature dependent indirect band gap value of BAs with and without spin-orbit coupling as a consequence of thermal expansion.}
	\label{fig:thermal-expansion}
\end{figure}

Using the quasiharmonic approximation \cite{Dove2005}, we consider lattice parameters ranging from $4.82$\,$\mathrm{\AA}$ to $4.86$\,$\mathrm{\AA}$ in eight equidistant steps. For each of these structures we calculate the Helmholtz free energy within the harmonic approximation to lattice dynamics \cite{Maradudin1963_harmonic_approximation}. 
Fig.\,\ref{fig:thermal-expansion_gap_a} depicts the calculated Helmholtz free energy relative to the static lattice energy of BAs as a function of lattice parameter and for temperatures ranging from $0$\,K to  $400$\,K. The grey points indicate the interpolated equilibrium lattice parameter and provide the quasiharmonic estimate of the equilibrium volume at the given temperature. The depicted temperature dependent minima show that the zero point quantum motion has the biggest effect on the lattice parameter, which increases by $0.01$\,\AA\@ compared to the static lattice value. The temperature induced lattice parameter change is smaller at only $0.006$\,\AA\@  in going between $0$\,K and $400$\,K. We estimate a room temperature thermal expansion coefficient of $4.6\times 10^{-6}$\,K$^{-1}$ which is in good agreement with the recently measured room temperature thermal expansion coefficient of $4.2\pm0.4\times10^{-6}$\,K$^{-1}$ \cite{Shi2019_BAs_thex}. In Fig. \ref{fig:thermal-expansion_gap_b} the thermal expansion driven temperature dependent indirect band gap with and without consideration of spin-orbit coupling is depicted. The graph shows that the indirect band gap increases by only $11$\,meV between $0$\,K and $400$\,K, an opposite trend to that exhibited by the changes driven by electron-phonon coupling, and with a magnitude about a tenth smaller. In addition, we calculate the effect of thermal expansion on the electron-phonon coupling strength at 300\,K. We use the equilibrium lattice parameter calculated within the quasiharmonic approximation at $300$\,K and evaluate the indirect band gap using the Monte Carlo sampling and the PBE functional for a $4\times4\times4$ supercell. The band gap correction has a value of $-151\pm5$\,meV which lies within the statistical uncertainty of the calculated value of $-142\pm6$\,meV without consideration of volume expansion. Therefore, we will only account for the effects of electron-phonon coupling when studying finite temperature optical absorption in Sec.\,\ref{optics} below.



\section{Phonon-mediated optical absorption} \label{optics}

\subsection{Computational details}

The finite temperature optical absorption of BAs is calculated using a similar approach to that described in Sec.\,\ref{sec:comp_det_elec} for the finite temperature electronic eigenvalues. The optical properties of semiconductors and metals such as absorption, reflectance, or energy loss spectra, are all determined by the frequency dependent complex dielectric function $\varepsilon_1(\omega) + i\varepsilon_2(\omega)$. We evaluate the imaginary part of the dielectric function within the dipole approximation using {\sc vasp}. For the calculation of the dielectric properties we reduce the energy cutoff to $250$\,eV (without loss of accuracy) and we sample the electronic BZ using a uniform $16\times16\times16$ $\mathbf{k}$-point grid and commensurate grids for the supercells. Energy conservation in the optical absorption process is enforced by a delta function smeared into a Gaussian function with a width of $80$\,meV.  We report results using the PBE functional in conjuction with a scissor operator chosen so as to reproduce the static lattice value of the band gap calculated at the HSE level of theory.

The finite temperature value of $\varepsilon_2$ is determined by replacing the electronic eigenvalue by $\varepsilon_2$ in Eq.\,(\ref{eq:elec_tdep}), an approach known as the Williams-Lax theory \cite{Williams1951_QZPE, Lax1952_Franck-condon_crystals, Giustino2015_EPC+optics2}. We then evaluate the corresponding integral using the stochastic Monte Carlo approximation combined with thermal lines \cite{Tomeu2016_TL} to accelerate the sampling.
We find that the finite temperature dielectric function converges using four sampling points in a $4\times4\times4$ supercell of BAs including $128$ atoms. 
After the calculation of $\varepsilon_2$, the real part of the dielectric function $\varepsilon_1$ is obtained using the Kramers-Kronig relation \cite{Kramers_Kronig_relation}, and the absorption coefficient is $\kappa(\omega)=\omega\varepsilon_2/ c n(\omega)$, where $c$ is the speed of light and $n(\omega)$ is the frequency dependent refractive index of the sample.

\subsection{Temperature dependent optical response}

\begin{figure}
	\centering
	\sidesubfloat[]{\includegraphics[scale=0.33]{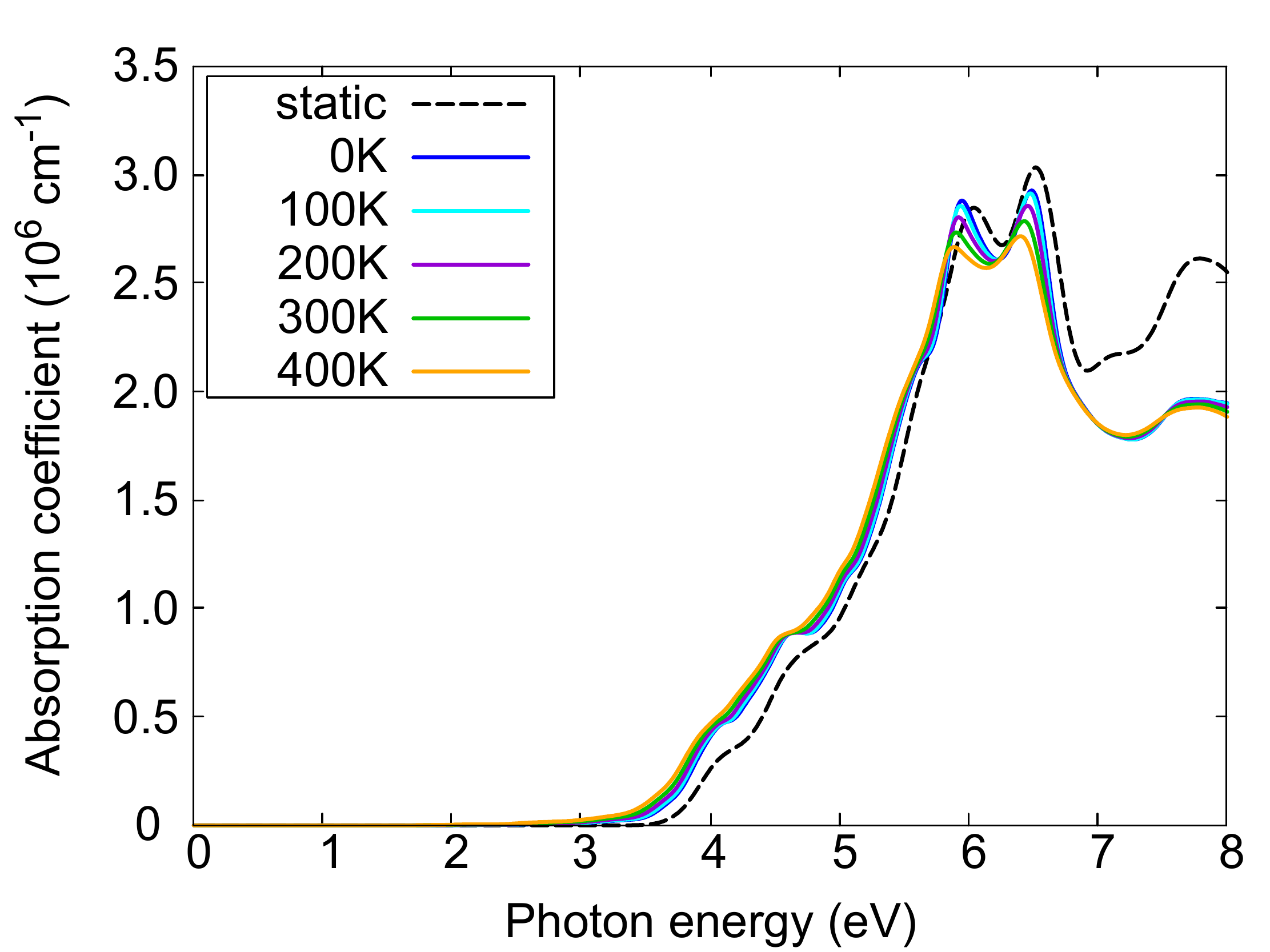}\label{fig:abs_spectra_a}}
	\hfil
	\sidesubfloat[]{\includegraphics[scale=0.33]{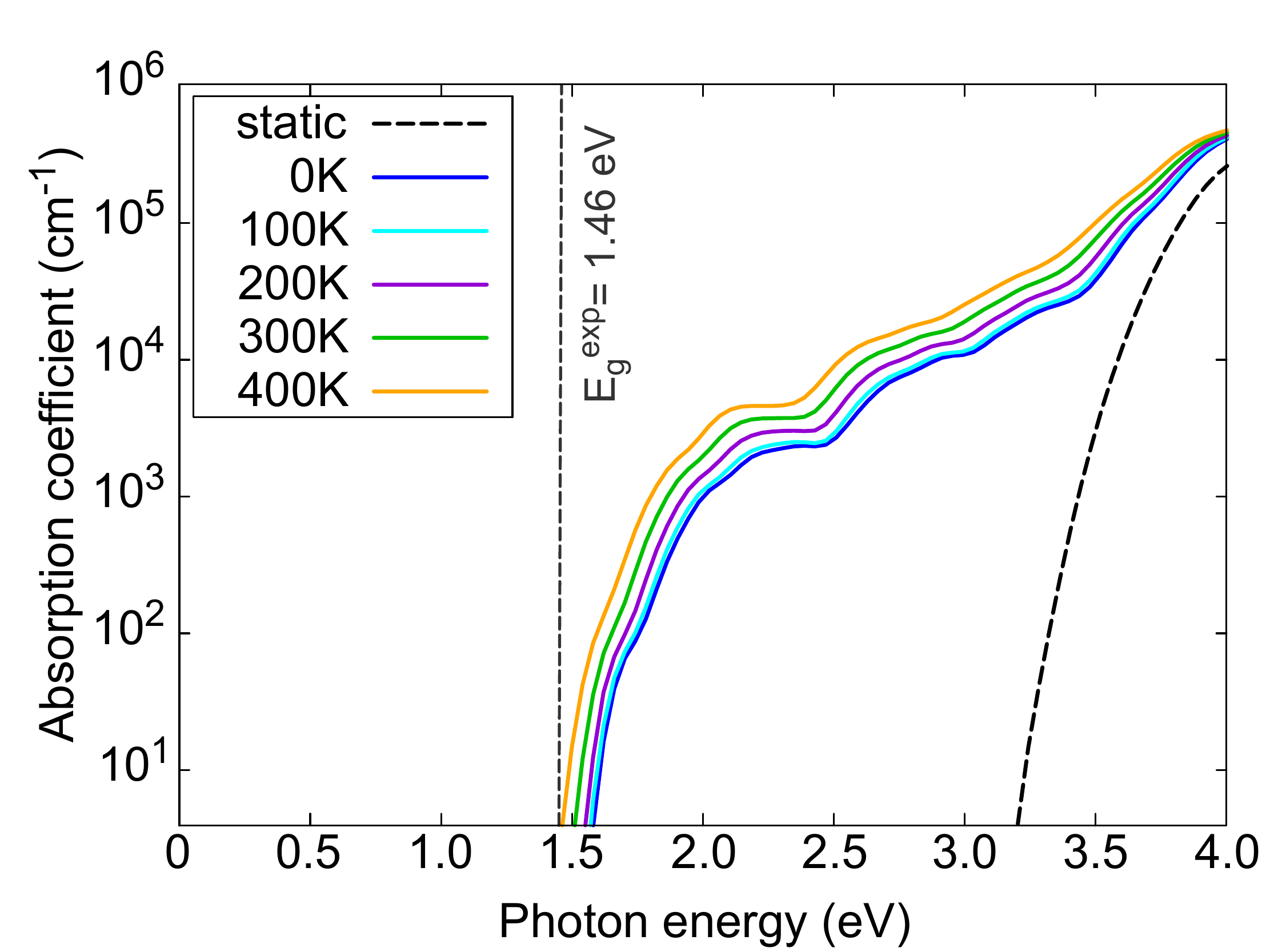}\label{fig:abs_spectra_b}}
	\caption{(a) Absorption coefficient of BAs using the PBE approximation to the exchange correlation functional. The calculations correspond to the stationary atoms at their equilibrium positions (dashed black line) and to the phonon-assisted contributions at temperatures varying from $0$\,K to $400$\,K (solid lines). We employ a scissor operator as discussed in the text and described in Ref.\,\cite{Giustino2015_EPC+optics2}. (b) Absorption onset of the absorption spectrum of BAs at the static lattice level (dashed black line) and at temperatures ranging from $0$\,K to $400$\,K (solid lines). The vertical dashed line depicts the experimental indirect band gap measured at room temperature.}
	\label{fig:abs_spectra}
\end{figure}

The effects of phonons on the optical properties of semiconductors are manifold, and here we focus on two different aspects of optical absorption in BAs. First, we consider phonon-assisted photon absorption across the indirect band gap from the valence band maximum to the conduction band mininum, a second order mechanism that conserves the overall energy and momentum throughout the absorption process. Second, we investigate the effect of increasing temperature on the spectrum: the absorption onset, the shape of the spectrum, and the absorption coefficient.

Figure \ref{fig:abs_spectra_a} shows the absorption spectrum of BAs in an energy range up to $8$\,eV, calculated with the PBE exchange-correlation functional in conjunction with a scissor operator obtained from the band gap difference between HSE and PBE (see Table \ref{tab:indirect_PBE_vs_HSE}). Figure \ref{fig:abs_spectra_b} shows the same spectrum in a logarithmic scale up to an energy of $4$\,eV.

The dashed black line of Fig.\,\ref{fig:abs_spectra} corresponds to the absorption spectrum of BAs calculated for the static lattice. 
The most significant difference between the static lattice level result and the spectrum at $0$\,K (solid blue line) is the position of the absorption onset which is clearest in the logarithmic plot of the absorption spectrum in Fig. \ref{fig:abs_spectra_b}. The absorption onset for the static lattice approximation occurs at $3.2$\,eV and corresponds to a transition across the direct optical band gap at $\Gamma$. The onset value of $3.2$\,eV is below the static lattice value of the band gap of $4.1$\,eV reported in Table \ref{tab:indirect_PBE_vs_HSE} due to the Gaussian smearing used for the calculation of $\varepsilon_2$, an effect that is magnified in the logarithmic scale. The onset is additionally biased due to the employed scissor operator which recovers the correction for the indirect band gap but still underestimates the effect on the optical band gap by 0.28\,eV. The absorption onset at $0$\,K is governed by zero point quantum motion, which enables the transition across the indirect band gap to occur with non-vanishing weight. As a consequence, the $0$\,K absorption onset is located over $1.5$\,eV below the static lattice absorption onset. This is a general feature of indirect band gap semiconductors, and can only be captured in optical response calculations when the effects of phonons are included. 

Increasing temperature leads to a red shift of the absorption onset, a result that is consistent with the findings of the temperature dependent band gap presented in Sec.\,\ref{sec:el-phonon-coupling}. We note that both the optical gap (Fig.\,\ref{fig:abs_spectra_a}) and the minimum gap (Fig.\,\ref{fig:abs_spectra_b}) are subject to this red shift with increasing temperature. Increasing temperature also leads to an increase of the absorption coefficient, best seen in the logarithmic plot in Fig.\,\ref{fig:abs_spectra_b}. Without phonons, only vertical transitions are allowed due to momentum conservation, but with phonons, transitions connecting different $\mathbf{k}$-vectors also have non-vanishing (second-order) transition matrix elements. This means that the absorption cross section increases with the phonon population due to the increase in available initial and final states, leading to an increase in the absorption coefficient. The magnitude of the increase is relatively low compared to the overall signal strength as the phonon-mediated opical absorption is a second order process, but appears to be a generic feature of indirect band gap semiconductors \cite{Noffsinger2012_Si_phonon,Giustino2015_EPC+optics2,Tomeu2018_BaSnO3}.

Another significant feature which arises in optical absorption spectra when including the interaction with the lattice is the smoothening of the absorption peaks with increasing temperature. With increasing number of phonons the accessibility of final states of different energies increases, leading to a broader energy spectrum and a smoother signal.

\section{Summary and Conclusions} \label{Summary}

In this work we have presented first principles calculations on the indirect band gap of BAs, taking into account the effects of temperature via the electron-phonon coupling and thermal expansion, the effects of electron correlation through exact exchange, and the role of the spin-orbit interaction. Electron-phonon coupling enables the phonon-mediated optical absorption across the minimum indirect band gap of BAs, a transition which within the standard static lattice approximation is prohibited due to momentum conservation. This results in an absorption onset which occurs at about $1.5$\,eV below the absorption onset calculated at the static lattice level. Furthermore, electron-phonon coupling also induces a red shift of the absorption onset and a smoothening of the absorption peaks with increasing temperature. We find that thermal expansion has a negligible effect on the optoelectronic properties of BAs.

Additionally, our results show that exact exchange included using the hybrid HSE functional enhances the electron-phonon driven band gap change with temperature by around $20$\%, indicating that electron-electron correlation significantly influences the electron-phonon matrix elements in BAs. In contrast, the spin-orbit interaction only modifies the electron-phonon coupling strength weakly.

The findings in our study underpin the potential of cubic boron arsenide as a material for use in optoelectronic devices. A comparison with silicon shows that, apart from the extraordinarily high thermal conductivity of BAs, which is six times higher than that of crystalline silicon, BAs exhibits a similar absorption onset to that of silicon \cite{Silicon-absorption} but has a higher absorption coefficient. The absorption probability of BAs is approximately four times higher in the region of the absorption onset than that of silicon. This feature leads to a higher absorption cross-section and potentially to a higher performance in operating devices compared to the silicon based analogues. 


Our work shows that there are manifold contributions that may play a significant role in determining the temperature dependence of the optical properties and ultimately the optoelectronic performance of semiconductors. These include electron-phonon coupling, thermal expansion, exact exchange, and spin-orbit coupling. These effects might also influence the extraordinary thermal conductivity properties of BAs, and our work demonstrates that a full characterization of this material requires careful validation of the level of theory used. We think that all of these effects should be considered when predicting novel materials for potential optoelectronic devices, and that this will become particularly important with increasing material complexity where an \textit{a priori} estimation of their relative strength might prove impossible.

\acknowledgements

The authors acknowledge support from the Winton Programme for the Physics of Sustainability, and B.M. also acknowledges support from Robinson College, Cambridge, and the Cambridge Philosophical Society for a Henslow Research Fellowship. Part of the calculations were performed using resources provided by the Cambridge Tier-2 system operated by the University of Cambridge Research Computing Service (http://www.hpc.cam.ac.uk) funded by EPSRC Tier-2 capital grant EP/P020259/1.

\bibliography{reference_BAs}

\end{document}